\documentclass[11pt]{article}

\usepackage[table]{xcolor}
\usepackage{graphicx} %
\usepackage{tikz}
\usepackage{indentfirst} %
\usepackage{setspace} %
\usepackage[longnamesfirst]{natbib} %
\usepackage{booktabs} %
\usepackage{rotating} %
\usepackage{amsmath} %
\usepackage[margin = 1in]{geometry}
\usepackage{subfig}
\usepackage{mdwlist}
\usepackage[hyphens]{url}
\usepackage{color}
\usepackage{verbatim}
\usepackage{multirow}
\usepackage{makeidx}  %
\usepackage{ifpdf}
\usepackage{url}
\usepackage{graphicx}
\usepackage{amssymb}
\usepackage{amsthm}
\usepackage{setspace}
\usepackage{natbib}
\usepackage{appendix}

\graphicspath{{./figs/}}

\newtheorem{proposition}{Proposition}

\newcommand{\EE}{\mathbb{E}}
\newcommand{\Var}{\mathrm{Var}}
\newcommand{\N}{\mathrm{N}}

\title{\vspace{-1cm} Algorithmic Decision Making\\in the Presence of Unmeasured Confounding\\\vspace{1cm}}
\author{%
	\makebox[.3\linewidth]{Jongbin Jung}\\Stanford University
	\and \makebox[.3\linewidth]{Ravi Shroff}\\New York University
    \and \makebox[.3\linewidth]{Avi Feller}\\UC-Berkeley
	\and \makebox[.3\linewidth]{Sharad Goel}\\Stanford University
}
\date{}

\begin{document}
\maketitle
\thispagestyle{empty}

\newcommand\independent{\protect\mathpalette{\protect\independenT}{\perp}}
\def\independenT#1#2{\mathrel{\rlap{$#1#2$}\mkern2mu{#1#2}}}

\begin{abstract}
On a variety of  complex decision-making tasks,
from doctors prescribing treatment to judges setting bail,
machine learning algorithms have been shown to
outperform expert human judgments.
One complication, however, is that
it is often difficult to anticipate the effects of
algorithmic policies prior to deployment,
making the decision to adopt them risky.
In particular, one generally cannot use historical data to directly observe
what would have happened had the actions recommended by the algorithm been taken.
One standard strategy is to model potential outcomes for alternative decisions
assuming that there are no unmeasured confounders (i.e., to assume ignorability).
But if this ignorability assumption is violated,
the predicted and actual effects of an algorithmic policy can diverge sharply.
In this paper we present a flexible, Bayesian approach to gauge the
sensitivity of predicted policy outcomes
to unmeasured confounders.
We show that this policy evaluation problem is a generalization of estimating heterogeneous treatment effects in observational studies, and so our methods can immediately be applied to that setting.
Finally, we show, both theoretically and empirically, that under certain conditions
it is possible to construct
near-optimal algorithmic policies even when
ignorability is violated.
We demonstrate the efficacy of our methods on a large dataset of judicial actions, in which one must decide whether defendants awaiting trial should be required to pay bail or can be released without payment.
\end{abstract}

\textbf{Keywords:} Bayesian sensitivity analysis, heterogeneous treatment effects, policy evaluation

\section{Introduction}

Machine learning algorithms are increasingly used by
employers, judges, lenders, and other experts to
guide high-stakes decisions.
These algorithms, for example, can be used to help
determine which job candidates are interviewed,
which defendants are required to pay bail,
and which loan applicants are granted credit.
At a high level, creating such algorithmic policies is straightforward.
First, based on detailed individual-level data on past decisions and outcomes,
one trains a statistical model to estimate the likelihood of key events under
the possible actions one could take.
For example, based on credit history,
one might estimate an applicant's risk of default if a loan is granted;
or based on criminal history,
one might estimate a defendant's likelihood of appearing at trial if
bail is set or, alternatively, if the defendant is released without requiring payment.
Second, with these individual-level estimates in hand,
one codifies a decision rule that balances competing interests.
For example, one might grant loans only to those deemed most creditworthy,
subject to limits on available funds;
or one might require bail only from defendants at greatest risk of
skipping trial if released, setting the risk threshold to balance flight risk
against the social and financial burdens of bail on defendants.

This basic strategy rests on the ability to accurately estimate counterfactual outcomes.
In reality, however, such predictions can be difficult.
For example, in the judicial context
we only observe whether or not a particular defendant failed to appear at trial
given the action the judge actually took
(i.e., requiring bail or not);
we do not observe what would have happened under the alternative judicial action.
If judges have information about defendants that is not recorded in the data,
and if this information correlates both with a judge's decision and a defendant's flight risk,
estimates of risk will in general be biased.
In statistical terms, there may be unmeasured confounding,
and, as a result, we cannot ignore the process by which individuals are assigned to treatment.

This fundamental limitation has two immediate consequences.
First, without accurate estimates of potential outcomes,
one cannot in general construct optimal algorithmic policies.
Second, and perhaps even more importantly,
one cannot precisely forecast the effects of deploying a given policy.
In particular, poor counterfactual estimates may lead lenders to take on
unexpectedly high levels of credit risk,
or courts to take on unexpectedly high levels of flight risk.
For organizations to accurately evaluate the implications of deploying decision algorithms,
it is thus important to gauge the sensitivity of forecasted effects to unmeasured confounding.

We make three key contributions in this article.
First, we develop a flexible Bayesian method to evaluate decision algorithms in the presence of unmeasured confounding.
Second, we show that the policy evaluation problem we consider here is
a generalization of estimating heterogeneous treatment effects in observational
studies---a problem of independent interest---and adapt our sensitivity analysis approach to that question.
Third, we show that in certain common scenarios,
one can construct near-optimal decision algorithms even if there is unmeasured confounding.
In particular, it is often possible to accurately rank individuals by risk---for example,
risk of default or risk of skipping trial---even if one cannot perfectly estimate
the true risk level itself.

To illustrate our methods, we consider judicial bail decisions,
starting with a detailed dataset of over 150,000 judgments in a large, urban jurisdiction.
Based on this information, we create realistic synthetic datasets for which
potential outcomes and selection mechanisms are, by construction, fully known.
We apply our methods to these synthetic datasets, selectively masking potential outcomes and
the data-generation process to mimic real-world applications.
Under a variety of conditions, we find that our approach accurately recovers
true policy effects and subgroup-level treatment effects,
despite the presence of unmeasured confounding.
In these simulations, our approach outperforms classical,
non-Bayesian sensitivity analysis techniques that recently have been adapted to the algorithmic
decision-making context~\citep{rosenbaum1983assessing,simplerules}.
When applied to the special case of estimating average treatment effects,
our general method compares favorably to existing,
state-of-the-art approaches that are tailored to that specific problem.
Finally, we show that on these synthetic  datasets,
we can construct near-optimal decision algorithms in the presence of unmeasured confounding.

\subsection{Related work}
\label{sec:related_work}

Our work touches on three related but distinct lines of research---algorithmic decision making, offline policy evaluation, and sensitivity analysis---which we briefly discuss here.
In field settings, decision makers often choose a course of action based on
experience and intuition rather than on statistical
analysis~\citep{klein2017sources}.  This includes doctors classifying patients
based on their symptoms~\citep{McDonald1996}, judges setting bail
amounts~\citep{Dhami2003} or making parole decisions~\citep{danziger_2011}, and
managers determining which ventures will
succeed~\citep{aastebro2006effectiveness} or which customers to
target~\citep{wubben2008instant}.
Despite the prevalence of this approach, a large body of work shows that in many
domains intuitive inferences are inferior to those based on statistical models~\citep{meehl1954clinical,dawes1979robust,dawes1989clinical,
camerer_johnson_1997,tetlock_2005,kleinberg_2017}. As such, algorithmic decision aids are now being used in diverse settings, including law enforcement, education, employment, and medicine~\citep{barocas2016big,berk2012criminal,chouldechova18,goel2016-risky,goel2016precinct,rudin_2017}.
In particular, applications of risk assessment algorithms have a long history in criminal justice,
dating back to parole decisions in the 1920s~\citep{burgess_1928}.
Several empirical studies have measured
the effects of adopting such decision aids.
For example, in a randomized controlled trial,
the Philadelphia Adult Probation and Parole
Department evaluated the effectiveness of a risk
assessment tool developed by \citet{berk2009},
and found the tool
reduced the burden on parolees without significantly
increasing rates of reoffense~\citep{ahlman2009}.

When determining whether to adopt a decision algorithm, it is important to anticipate its likely effects using only historical data, as ex-post evaluation may be expensive or otherwise impractical. This general problem is known as
offline policy evaluation in the machine learning community.
One approach to this problem is to first assume that treatment assignment is ignorable given the observed covariates,
after which a variety of modeling techniques are theoretically justified, including regression, matching, and
doubly robust estimation~\citep{dudik_2011,zhang2012robust,zhang2015using,athey2017}.
A related issue---sometimes referred to as optimal treatment assignment---is to select an optimal policy for the task at hand, given a method of evaluation and a set of candidate policies~\citep{gail1985testing,manski2004statistical,dehejia2005program,chamberlain2011bayesian}.

As the assumption of ignorability is often unrealistic in practice, sensitivity analysis methods seek to gauge the effects of unmeasured confounding on predicted outcomes.
The literature on sensitivity analysis dates back at least to the work of \citet{cornfield1959smoking} on the link between smoking and lung cancer.
In a seminal paper,~\citet{rosenbaum1983assessing} propose a framework for sensitivity analysis in which a binary unmeasured confounder affects both a binary
response variable and a binary treatment decision.
By sweeping over various plausible values for the strength of the assumed relationships, one obtains a corresponding range of estimated effects.
Some extensions of the Rosenbaum and Rubin approach
to sensitivity analysis
include allowing for non-binary response variables,
and incorporating machine learning methods into the estimation process~\citep{imbens2003,carnegie2016assessing, dorie2016flexible}.
A complementary line of work extends classical sensitivity analysis by taking a Bayesian perspective and averaging, rather than sweeping, over values of the sensitivity parameters~\citep{mccandless2007bayesian,mccandless2017comparison}.
In this setup,
even a weakly informative prior distribution over the sensitivity parameters can exclude estimates that are not supported by the data.
\citet{simplerules} have recently extended the Rosenbaum and Rubin sensitivity framework to offline policy evaluation, though
we do not know of any existing Bayesian approaches to the problem.

\section{Bayesian sensitivity analysis for policy evaluation}
\label{sec:policy_eval}

\subsection{Preliminaries and problem definition}
\label{ssec:problem_def}

A common goal of causal inference is to estimate the average effect of a binary treatment $T \in \{0,1\}$ on a response $Y \in \mathbb{R}$.
That is, one often seeks to estimate $\EE[Y(1) - Y(0)]$,
where $Y(0)$ and $Y(1)$ denote the potential outcomes under the two possible treatment values.
Here we consider a variant of this problem that arises in many applied settings.
Namely, given a policy $\pi: \mathbb{R}^m \mapsto \{0,1\}$ that assigns treatment based on individual characteristics $X \in \mathbb{R}^m$,
we seek to estimate the average response $V^{\pi} = \EE[Y(\pi(X))]$.
In our judicial bail application (described in detail in Section~\ref{sec:judicial}),
$\pi$ is an algorithmic rule that determines which
defendants are released on their own recognizance (RoR) and which are required to pay bail,
$Y \in \{0, 1\}$ indicates whether a defendant
appears at trial,
$X$ is a vector of observed covariates, and
$V^{\pi}$ is the expected proportion of defendants who fail to appear in court when the rule is followed.

Although $V^\pi$ is not a classical causal effect, it is closely related to several standard estimands in causal inference.
To see this, first define the subgroup average treatment effect for a group $G \subseteq \mathbb{R}^m$ to be $\EE[Y(1) - Y(0) \mid X \in G]$.
Next,
define the policy $\pi_G$
as \begin{equation*}
\pi_G = \left\{
\begin{array}{rl}
1 & \text{if } X \in G\\
0 & \text{if } X \notin G.\\
\end{array} \right.
\end{equation*}
In particular, $\pi_{\emptyset} \equiv 0$.
Let $\mathbf{I}(\cdot)$ be an indicator function evaluating to 1 if its argument is true and to 0 otherwise.
Then,
\begin{align*}
\EE[Y(1) - Y(0) \mid X \in G] & =
\frac{\EE[ (Y(1) - Y(0))\mathbf{I}(X \in G)]}{\Pr(X \in G)} \\
&= \frac{\EE[ Y(1)\mathbf{I}(X \in G) + Y(0)\mathbf{I}(X \notin G) - Y(0)]}{\Pr(X \in G)} \\
&= \frac{\EE[ Y(\pi_G(X)) - Y(\pi_\emptyset(X))]}{\Pr(X \in G)} \\
&= \frac{V^{\pi_G} - V^{\pi_{\emptyset}}}{\Pr(X \in G)}.
\end{align*}
Hence, if one can estimate $V^{\pi}$ for arbitrary policies $\pi$,
then one can estimate subgroup average treatment effects for arbitrary groups $G$.
The case $G = \{x\}$ corresponds to the conditional average treatment effect, and the case $G = \mathbb{R}^m$ corresponds
to the average treatment effect.
We can thus think of policy evaluation as a generalization of estimating heterogeneous treatment effects.

\subsection{Policy evaluation without unmeasured confounding}
\label{ssec:eval_ignorability}

To motivate the standard approach for estimating $V^{\pi} = \EE[Y(\pi(X))]$,
we decompose it into the sum of two terms:
\begin{align*}
V^{\pi} %
		&= \EE[Y(T) \mid \pi(X) = T] \cdot \Pr[\pi(X) = T] +
        \EE[Y(1-T) \mid \pi(X) \neq T] \cdot \Pr[\pi(X) \neq T].
\end{align*}
The proposed and observed policies are the same for the first term but differ for the second term.
Thus estimating the first term is straightforward, since we directly observe the outcomes for this group, $Y^{\text{obs}} = Y(T)$. The key statistical difficulty is estimating the second term, for which we must impute the missing potential outcomes, $Y^{\text{mis}} = Y(1-T)$. That is, in these cases we must estimate what would have happened had the recommendation been followed, rather than the observed action. This challenge is known as the fundamental problem of causal inference~\citep{holland1986statistics}.

To estimate $V^{\pi}$,
it is common to start with
a sample of historical data $\{(x_i, t_i, y_i(t_i))\}_{i=1}^n$ on individual covariates, treatment decisions, and observed outcomes.
We can then estimate $V^{\pi}$ via:
\begin{align}
\label{eq:v_pi_decomp}
\hat{V}^{\pi}
& = \frac{1}{n} \left[ \sum_{\pi(x_i) = t_i}
  y_i^\text{obs} + \sum_{\pi(x_i) \neq t_i} \hat{y}_i^\text{mis} \right],
\end{align}
where $y_i^{\text{obs}}$ is the observed outcome for individual $i$ when the proposed and observed policies agree.
A simple approach to estimating the unobserved potential outcomes
$\hat{y}_i^{\text{mis}}$ in the second term is
to directly model the outcomes conditional on treatment and observed covariates, sometimes referred to as response surface modeling~\citep[e.g.,][]{hill2012}.
In the case of pretrial detention, we can fit logistic regressions that estimates a defendant's likelihood of failing to appear at court when either released (RoR) or detained (set bail), conditional on the available information.
The unobserved potential outcome is then imputed via the corresponding fitted model.

This direct modeling approach implicitly assumes that the treatment is \emph{ignorable} given the observed covariates (i.e., that there is no unmeasured confounding).
Formally, ignorability means that
\begin{equation}
Y(0), Y(1) \independent T \mid  X.
\end{equation}
In other words, conditional on the observed covariates, those who receive treatment are similar to those who do not.
In the pretrial context, ignorability means that conditional on
the observed covariates, those who are RoR'd are similar to those who are not. In particular, this assumption excludes the possibility of
judges having access to information not recorded in the data---such as a defendant's demeanor---that affects both the judge's decision and the defendant's likelihood of appearing at trial.
In many situations, ignorability is a strong assumption.
The main contribution of this paper is a Bayesian approach to assessing the sensitivity of policy outcomes $V^\pi$ to violations of ignorability, which we present next.

\subsection{Policy evaluation with unmeasured confounding}
\label{ssec:bsa}
When ignorability does not hold, the resulting estimates of $V^\pi$ can be strongly biased.
To address this issue,
the sensitivity literature typically starts by assuming that there is an unmeasured confounder $U \in \mathbb{R}$ for each individual such that
ignorability holds given both $X$ and $U$:
\begin{equation}
\label{eq:ignorability_sens}
Y(0), Y(1) \independent T \mid X, U.
\end{equation}
In the pretrial setting, $U$ is meant to summarize all relevant information observed by a judge when making a decision but not recorded in the available data.
One then generally imposes additional structural assumptions on the form of $U$ and its relationship to decisions and outcomes.
We follow this basic template to obtain sensitivity estimates for policy outcomes.

At a high level, we model the observed data $\{(x_i, t_i, y_i(t_i))\}_{i=1}^n$ as draws from parametric distributions depending on the measured covariates $x_i$ and the unmeasured, latent covariates $u_i$:
\begin{align*}
y_i(0) & \sim f(x_i, u_i; \alpha) \\
y_i(1) & \sim g(x_i, u_i; \beta) \\
t_i & \sim h(x_i, u_i; \gamma) \\
y_i & = y_i(t_i)
\end{align*}
where $\alpha$, $\beta$, $\gamma$, and $u_i$
are latent parameters with weakly informative priors.
The inferred joint posterior distribution on these parameters then yields estimates of $y_i(0)$ and $y_i(1)$ for each individual $i$.
Finally, the posterior estimates of the potential outcomes yield a posterior estimate of $V^{\pi}$ via Eq.~\eqref{eq:v_pi_decomp}
that accounts for unmeasured confounding.

Our strategy, while straightforward, differs in three important ways from classical sensitivity methods.
First, we directly model the treatment and potential outcomes, allowing for flexibility in the functional forms of $f$, $g$ and $h$.
This flexibility is enabled by recent computational advances in black-box inference for generative models~\citep{stan};
in the past, it was often necessary to analytically derive and maximize the corresponding likelihood function, significantly constraining model structure.
Second, our Bayesian approach
automatically excludes values of the latent variables that are not supported by the data~\citep{mccandless2017comparison}.
In contrast, previous methods generally require more extensive parameter tuning to obtain reasonable estimates.
Finally, as noted earlier, our focus is on the more general problem of estimating policy outcomes rather than the average treatment effects considered in past work.

We now describe the specific forms of $f$, $g$, and $h$ that we use throughout our analysis.
First, we reduce the dimensionality of the observed covariate vectors $x$  down to three key quantities:
\begin{enumerate}
\item $\mu_0(x) = \EE[Y(0)\mid X=x]$, the conditional average response if $T = 0$;
\item $\mu_1(x) = \EE[Y(1)\mid X=x]$, the conditional average response if $T = 1$; and
\item $e(x) = \Pr(T = 1\mid X=x)$, the propensity score.
\end{enumerate}
In practice, these three quantities can be estimated
via standard prediction models, like regularized regression.

Next, we divide the data into $K$ approximately equally sized groups, ranking and binning by the estimated outcome $\hat{\mu}_0$.
(One might alternatively group the data by $\hat{\mu}_1$, or even jointly by $\hat{\mu}_0$ and $\hat{\mu}_1$.)
Denote the group membership of observation $i$ by $k[i] \in \{1, 2, \dots, K\}$.
Then we model the observed data as follows:
\begin{align*}
y_i(0) & \sim \textrm{Bernoulli} \left(\textrm{logit}^{-1}\big(\alpha_{0,k[i]} + \alpha_{\hat{\mu}_0,k[i]}\hat{\mu}_0(x_i) + \alpha_{u,k[i]}u_i\big) \right) \\
y_i(1) & \sim \textrm{Bernoulli} \left(\textrm{logit}^{-1}\big(\beta_{0,k[i]} + \beta_{\hat{\mu}_1,k[i]}\hat{\mu}_1(x_i) +
  \beta_{u,k[i]}u_i) \right)\\
t_i & \sim \textrm{Bernoulli} \left(\textrm{logit}^{-1}(\gamma_{0,k[i]} + \gamma_{\hat{e},k[i]}\hat{e}(x_i) + \gamma_{u,k[i]}u_i) \right)\\
y_i & = y_i(t_i).
\end{align*}
In each of the first three equations above,
outcomes are modeled as draws from a Bernoulli distribution whose mean depends on both
the reduced covariates---$\hat{\mu}_0(x)$, $\hat{\mu}_1(x)$, and $\hat{e}(x)$---and the unmeasured confounder $u$.
To support complex relationships between the predictors and outcomes, we allow the coefficients to vary by group $k[i]$.
That is, we model mean response as piecewise linear on the logit scale.
Finally, to complete the Bayesian model specification, we must describe the prior distribution on the parameters.
We provide these details in the Appendix,
where we also show that our results are largely invariant
to the exact choice of priors.

\section{An application to judicial decision making}
\label{sec:judicial}

To demonstrate our general approach to policy evaluation,
we now consider in detail the case of algorithms designed to aid judicial decisions~\citep{simplerules, kleinberg_2017}.
In the U.S. court system, pretrial release determinations are among the most common and consequential decisions for criminal defendants. After a defendant is arrested, he is usually arraigned in court,
where a prosecutor presents a written list of charges.
If the case proceeds to trial, a judge must decide whether the defendant should be released
on his own recognizance (RoR) or subject to money bail, where release is conditional on providing
collateral meant to ensure appearance at trial.
Defendants who are not RoR'd and who cannot post bail themselves may await trial in jail,
or pay steep fees to a bail bondsman to post bail on their behalf.
Judges must therefore balance the burden that setting bail places on a defendant
against the risk that the defendant may fail to appear (FTA) for his trial.\footnote{In many jurisdictions, judges may also consider the risk that a defendant will commit a new crime
if released when deciding whether or not to set bail, but not in the jurisdiction we consider.}

Here we consider algorithmic policies for assisting these judicial decisions, recommending either RoR or bail
based on recorded characteristics of a defendant and his case. The policy evaluation problem is to estimate, based only on historical data, the proportion of defendants who would fail to appear if the algorithmic recommendations were followed.
As discussed above, this is statistically challenging
because one does not always observe what would have occurred  had the algorithmic policy been followed.
In particular, if the policy recommends releasing a defendant who was in reality detained, or recommends detaining a defendant who was in reality released, the relevant counterfactual is not observed.
Further, since judges may---and likely do---base their decisions in part on information that is not recorded in the data,
direct outcome models ignoring unmeasured confounding may be badly biased for counterfactual outcomes.
We thus allow for a real-valued unobserved covariate $u$
that affects both a judge's decision (RoR or bail) and also the outcome (FTA or not) conditional
on that decision.
For example, $u$ might correspond to a defendant's perceived demeanor,
with seemingly responsible defendants more likely to
be RoR'd and also more likely to appear at their court proceedings.
Our goal is to assess the sensitivity of flight risk  estimates to such unmeasured confounding.

\subsection{The effects of unmeasured confounding}
\label{ssec:real_data}

Our analysis is based on 165,055 adult cases involving nonviolent
offenses charged by a large urban prosecutor's office and arraigned
in criminal court between 2010 and 2015.
These cases do not include instances where defendants accepted a plea deal at arraignment,
where no pretrial release decision is necessary.
For each case, we have 49 features describing characteristics
of the current charges (e.g., theft, gun-related), and 15 features describing
characteristics of the defendant (e.g., gender, age, prior arrests). We
also observe whether the defendant was RoR'd, and whether he
failed to appear at any of his subsequent court dates. Applying the notation introduced in Section~\ref{sec:policy_eval} to this dataset, $x_i$ refers to a vector of all observed characteristics
of the $i$-th defendant, $t_{i} = 1$ if bail was set, and $y_i = 1$ if the defendant failed to appear at court.
Overall, 69\% of defendants in our dataset are RoR'd, and 15\% of RoR'd defendants fail to appear.
Of the remaining 31\% of defendants for whom bail is set,
9\% fail to appear.
As a result, the overall FTA rate is 13\%.
To carry out our analysis, we first randomly select 10,000 cases\footnote{We
  use a relatively small number of cases for the test set to mitigate issues of computational
  complexity.}
from the full dataset which we set aside as our final test data.
The remaining cases are split into two training data folds of equal size.

We begin by constructing a family of algorithmic decision rules, following the procedure of \citet{simplerules}.
On the first training fold of about 77,500 cases,
we fit an $L^1$-regularized (lasso) logistic regression model to estimate the probability of FTA given release, $\hat{\mu}_0^{\pi}(x_i)$
That is, we fit a logistic regression with the left-hand side indicating whether a defendant failed to appear at any of his court dates, and the right-hand side comprised of all available covariates
for the subset of defendants that the judge released.
We use the superscript $\pi$ to indicate that these estimates are computed on the first fold of data,
and are used exclusively to define the policies, not to evaluate them.
With these estimates in hand, we construct a family of polices $\{\pi_s\}$, indexed by a risk threshold $s$ for releasing individuals:
\[ \pi_s(x_i) = \begin{cases}
1 & \text{if } \hat{\mu}_0^{\pi}(x_i) > s\\
0 & \text{if } \hat{\mu}_0^{\pi}(x_i) \leq s.
\end{cases}\]
For example, the policy $\pi_{0.1}$ recommends that bail be set for a defendant if and only if his estimated flight risk if released, $\hat{\mu}_0^{\pi}(x_i)$, is at least 10\%.
These policies, which are based on a ranking of defendants by risk, are similar to pretrial algorithms used in practice~\citep{corbett-davies_2017}.

Given the family of policies $\{\pi_s\}$ defined above, we next estimate the proportion of defendants who would fail to appear under each policy, accounting for unmeasured confounding.
To do so, on the second training fold of about 77,500 cases, we fit three $L^1$-regularized logistic regression models to estimate each individual's likelihood of failing to appear if he were RoR'd or were required to pay bail---$\hat{\mu}_0(x_i)$ and $\hat{\mu}_1(x_i)$, respectively---as well as each individual's likelihood of having bail set, $\hat{e}(x_i)$.
Finally, on the test fold of data, consisting of 10,000 cases,
we fit our Bayesian sensitivity model of Section~\ref{ssec:bsa}, using the estimated quantities $\hat{\mu}_0(x_i)$, $\hat{\mu}_1(x_i)$, and $\hat{e}(x_i)$,
and setting the number of groups $K$ to 10.
Our use of a random walk prior on the coefficients
helps to ensure that our results are largely independent of
the specific value of $K$ (see the Appendix for details).

\begin{figure*}[t]
  \centering
  \includegraphics[width=6cm]{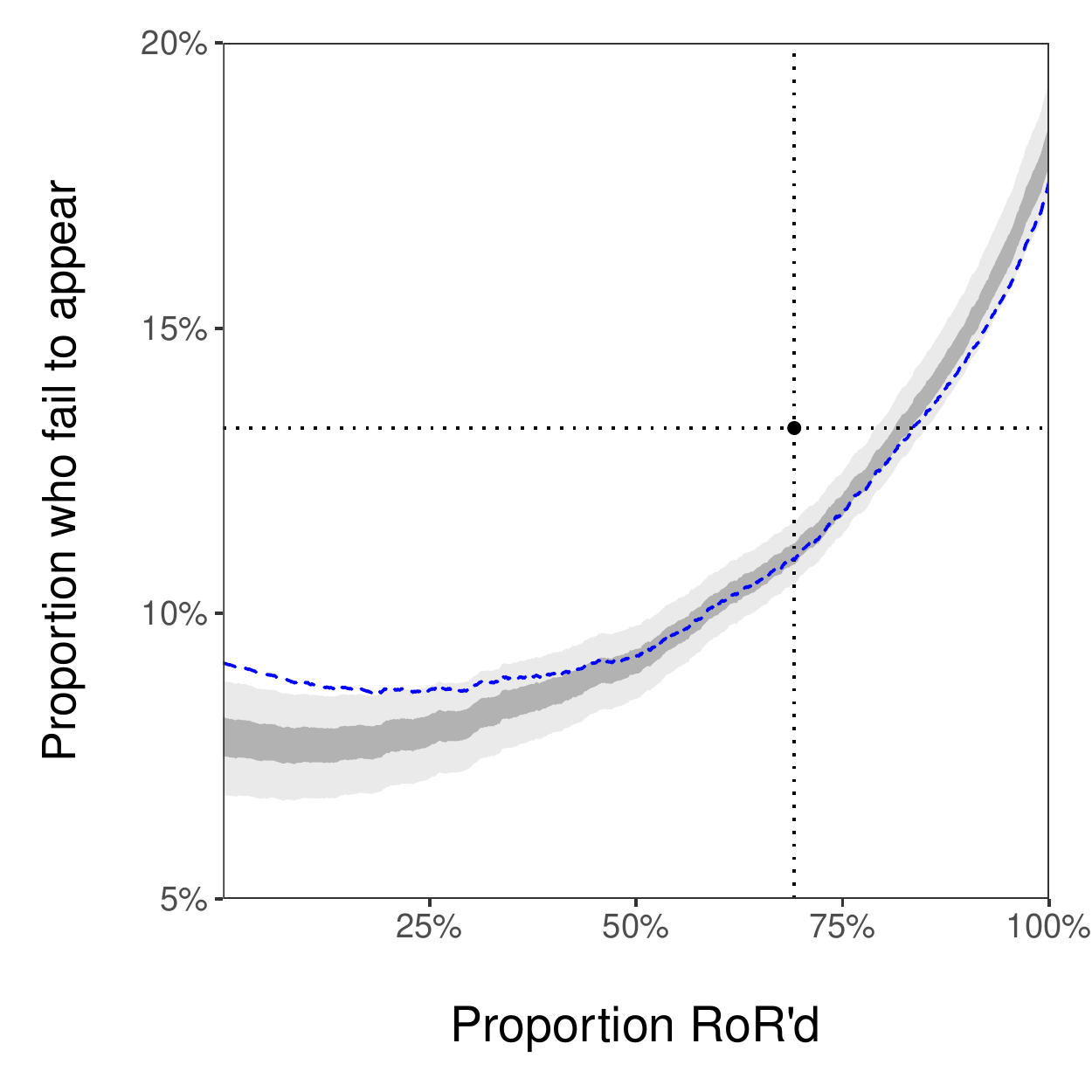}
  \caption{\emph{Sensitivity estimates for flight risk under a family of policies $\pi_s$. The blue line indicates estimates assuming no unmeasured confounding. The light gray band indicates a 95\% credible interval under our model of unmeasured confounding, and the dark gray band indicates a 50\% credible interval.}}
\label{fig:fulldata-sensitivity}
\end{figure*}

The results of this analysis are plotted in Figure~\ref{fig:fulldata-sensitivity}.
The blue dashed line shows, for each policy $\pi_s$, the proportion RoR'd and the estimated proportion that FTA under the policy, $\hat{V}^{\pi_s}$, estimated under no unobserved confounding.
For reference, the black point shows the status quo: judges in our dataset release 69\% of defendants, resulting in an
overall FTA rate of 13\%.
The light and dark gray bands show, respectively, the 95\% and 50\% credible intervals that result from the sensitivity analysis procedure under our model of unobserved confounding.

Figure~\ref{fig:fulldata-sensitivity} illustrates three key points.
First, for policies that RoR almost all defendants (toward the right-hand side of the plot), the blue line lies below our sensitivity bands, in line with expectations.
If there is unmeasured confounding, we would expect those who were in reality detained to be riskier than they seem from the observed data alone;
as a result, the direct outcome model
underestimates the proportion of defendants that would fail to appear if all (or almost all) such defendants are RoR'd.
Conversely, for policies that recommend bail for almost all defendants (toward the left-hand side of the plot), the blue line lies above the sensitivity bands,
as we would expect, because those who were in reality released are likely less risky than they appear from the observed data alone.
Second, as the policies move further from the status quo, toward the left- and right-hand extremes of the plot, the sensitivity bands grow in width, indicating greater uncertainty.
This pattern reflects the fact that the data provide less direct evidence of flight risk as polices diverge from the status quo,
heightening the potential impact of unmeasured confounding.
Finally, even after accounting for unmeasured confounding, there are algorithmic policies that do substantially better than the status quo, in the sense that they RoR more defendants and simultaneously achieve a lower overall FTA rate.

\begin{figure*}[t]
  \centering
  \includegraphics[width=14cm]{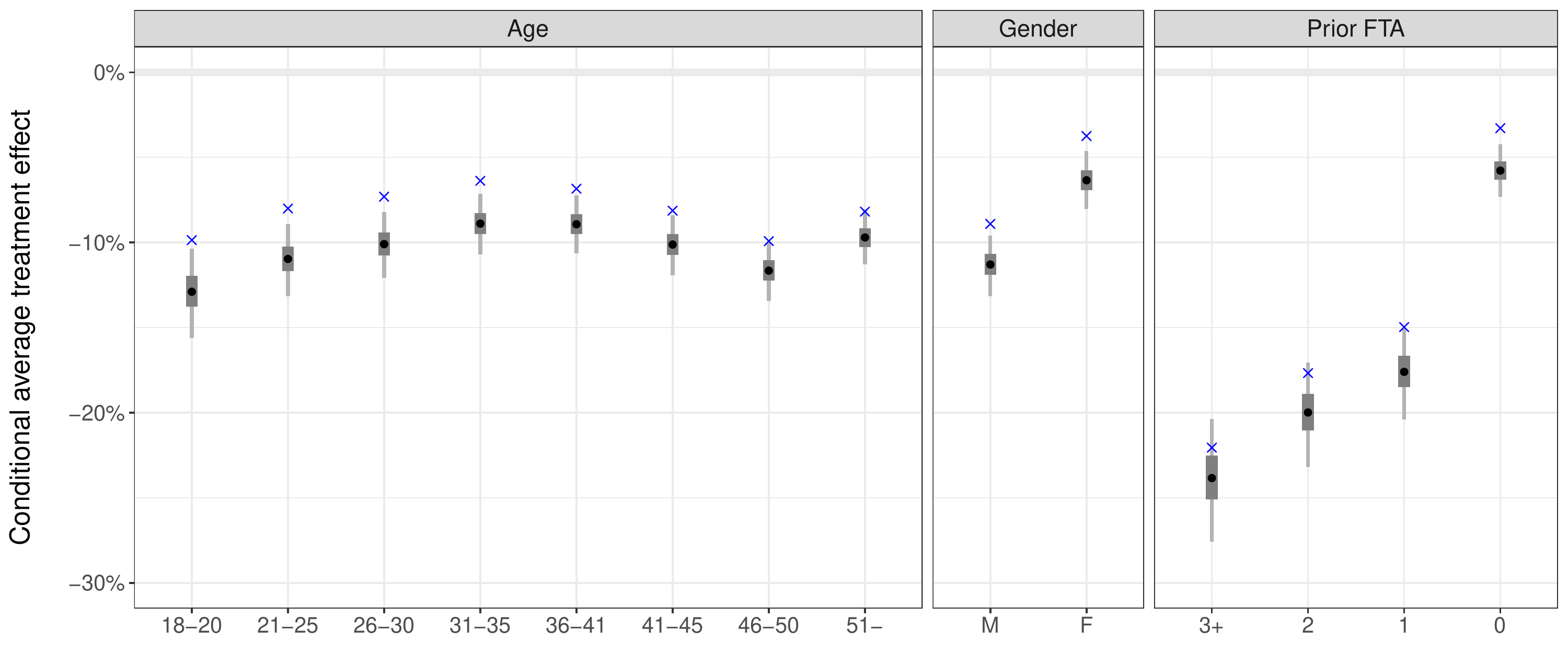}
  \caption{\emph{Heterogeneous treatment effects across groups defined by age, gender, and number of previously missed court appearances.
  The blue \texttt{X}'s
      mark the direct outcome model estimates ignoring unmeasured confounding, the gray lines indicate 50\% and 95\% credible
      intervals resulting from our sensitivity analysis, and the points indicate the median of the posterior distribution.}}
\label{fig:fulldata-het-tx}
\end{figure*}

As discussed above,
policy evaluation is a generalization of estimating heterogeneous treatment effects.
As such, our approach immediately
yields estimates of conditional average treatment effects under our model of unmeasured confounding.
For groups defined by age, gender, and
number of previously missed court appearances,
Figure~\ref{fig:fulldata-het-tx} displays
estimated effects of setting bail versus releasing all defendants in each group.
In each case,
the blue \texttt{X} is the estimated difference in FTA rate from the direct outcome model without unmeasured confounding.
The thick and thin gray bars indicate 50\% and 95\% credible intervals resulting from our sensitivity analysis,
and the point shows the median posterior subgroup treatment effect.
Across all subgroups, we can see that the estimates ignoring unmeasured confounding underestimate the magnitude of the effect (i.e., are closer to zero),
compared to estimates under our model of unobserved confounding.
The plot also shows that there is considerable  heterogeneity across defendants, particularly  when stratified by number of prior FTAs, highlighting the importance of considering conditional average treatment effects.
For example, for defendants with three or more previously missed court appearances, the average treatment effect is almost 20 percentage points lower than for defendants with no such lapses.

\subsection{A simulation study}
\label{sec:simulation}

Our results above quantify the sensitivity of policy outcomes under one particular model of unmeasured confounding.
Without further analysis, however, it is difficult to gauge whether that approach indeed provides accurate estimates of the true outcomes under such polices.
To address this question, we next evaluate our method on a series of realistic, synthetic datasets for which the true outcomes are, by construction, known.

\begin{figure*}[t]
  \centering
  \includegraphics[width=14cm]{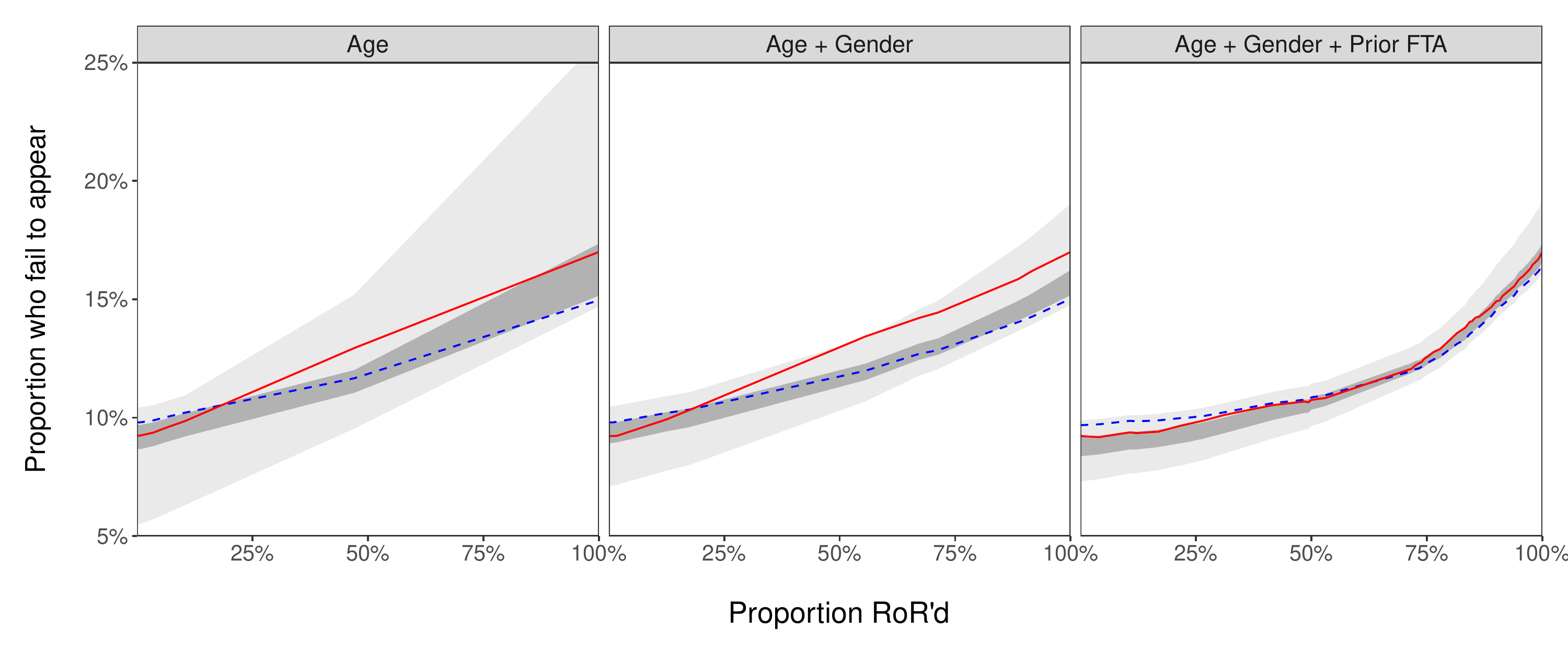}
  \caption{\emph{Sensitivity estimates on three synthetic datasets with varying degrees of unmeasured confounding.
    In each panel, datasets are restricted to contain only a subset of covariates: age (left);
    age and gender (center); age, gender, and prior FTAs (right).
    As in Figure~\ref{fig:fulldata-sensitivity},
the blue lines show estimates based on direct outcome models ignoring unmeasured confounding,
and the gray bands represent 50\% and
95\% credible intervals based on our sensitivity analysis.
The red lines show the true policy outcomes.
}}
\label{fig:synthetic-sensitivity}
\end{figure*}

We begin with the dataset introduced in Section~\ref{ssec:real_data},
a collection of 165,055 cases $\{(x_i, t_i, y_i(t_i))\}$,
where $x_i$ is a vector of covariates describing the defendant and
characteristics of the charge, $t_i = 1$ if the defendant was required to pay bail, and
$y_i(t_i) = 1$ if the defendant failed to appear for his court appearance.
To assess the performance of our sensitivity analysis procedure,
we create a synthetic dataset where \emph{both} potential outcomes for each defendant---$y_i(0)$ and $y_i(1)$---are known.
We do so by first estimating $\hat{\mu}_0(x_i)$, $\hat{\mu}_1(x_i)$,
and $\hat{e}(x_i)$ using $L^{1}$-regularized logistic regression, as above.
Then, for each individual in the original dataset, we
create synthetic potential outcomes and treatment assignments via
independent Bernoulli draws based on these estimated probabilities:
\begin{align*}
y_i'(0) & \sim \text{Bernoulli}(\hat{\mu}_0(x_i)) \\
y_i'(1) & \sim \text{Bernoulli}(\hat{\mu}_1(x_i)) \\
t_i' & \sim \text{Bernoulli}(\hat{e}(x_i)).
\end{align*}
We denote the resulting synthetic, uncensored dataset by
$\Omega = \{(x_i, t_i', y_i'(0), y_i'(1))\}$.
Because both potential outcomes are listed,
for any policy $\pi$ applied to $\Omega$
we can exactly calculate $V^{\pi}$,
the proportion of defendants that fail to appear for their trial.
We note that our synthetic dataset, by construction, satisfies ignorability.
That is, conditional on the covariates $x_i$,
treatment assignment is independent of the potential outcomes.

We next censor $\Omega$ in two ways.
First, we restrict the observed covariates to a subset $x_i'$.
In our simulation study, we consider three sets of restricted covariates:
age; age and gender; and age, gender, and prior number of missed court appearances.
Second, for each individual we remove the potential outcome for the action not taken, keeping only $y_i'(t_i')$.
Thus, for each of our three choices of $x_i'$, we have a
dataset of the form
$\Omega^{'} = \{(x_i', t_i', y_i'(t_i'))\}$.
The covariates not included in $x_i'$
correspond to unmeasured confounding.
Starting from these three synthetic datasets,
we carry out the policy construction and sensitivity analysis procedure described in Section~\ref{ssec:real_data}.
For example, in the case where $x_i'$ is comprised of age and gender, we first find release polices based on the available covariates (i.e., age and gender), and then run our Bayesian sensitivity analysis to estimate the effects of unmeasured confounding.

\begin{figure*}[t]
  \centering
  \includegraphics[width=14cm]{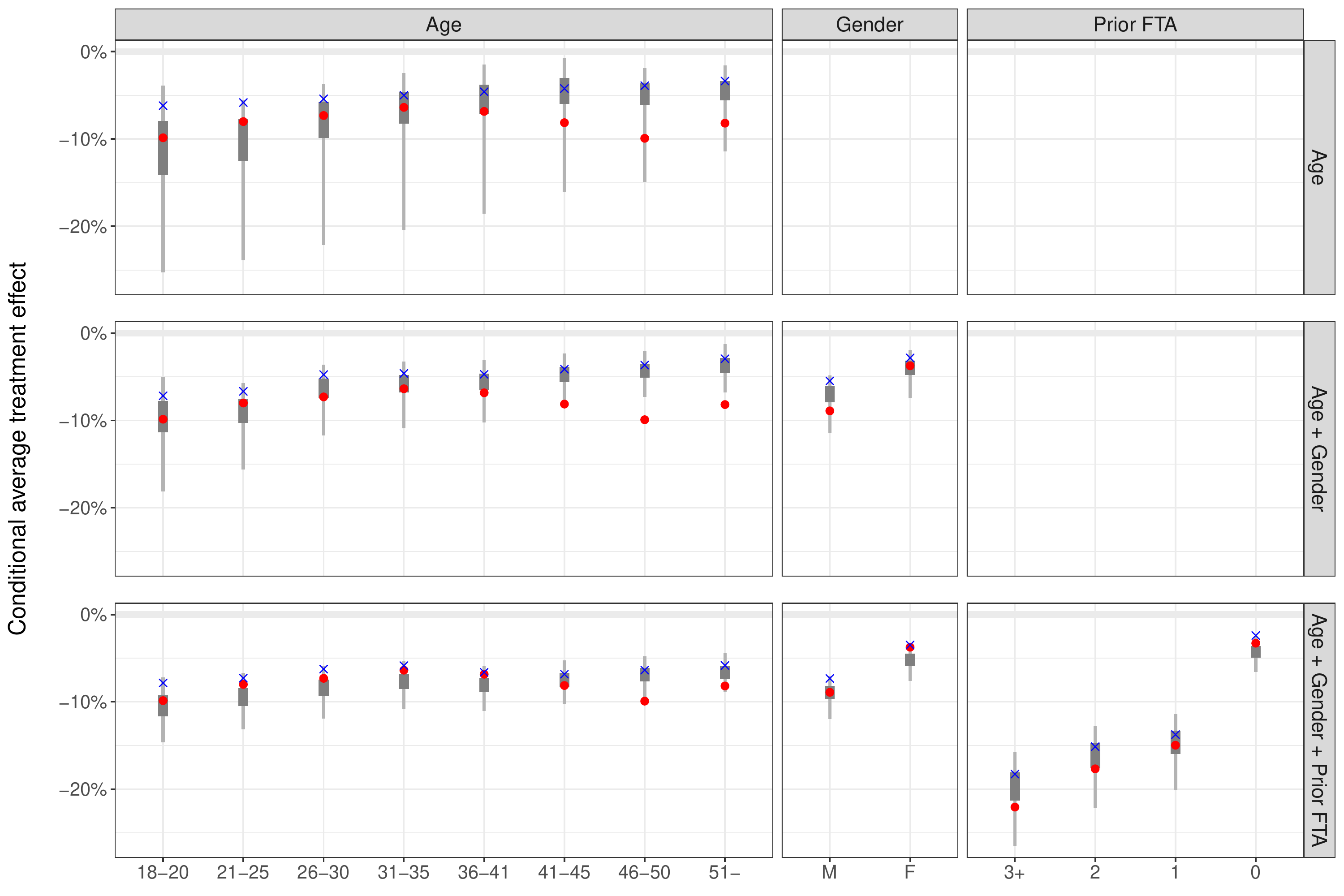}
  \caption{\emph{Estimates of subgroup treatment effects on three synthetic datasets with varying degrees of unmeasured confounding.
  Rows correspond to datasets comprised of age (top), age and gender (middle), and age, gender, and number of previously missed court appearances (bottom).
  In each panel, \texttt{X}'s indicate estimates from direct outcome models ignoring unmeasured confounding, the gray lines indicate 50\% and 95\% credible intervals from our sensitivity analysis, and the red points mark the ground truth treatment effects.}}
\label{fig:synthetic-het-tx}
\end{figure*}

The results of our simulation study are shown in Figure~\ref{fig:synthetic-sensitivity},
with each panel corresponding to a different choice of
$x_i'$ and thus a different degree of unmeasured confounding.
As in Figure~\ref{fig:fulldata-sensitivity},
the blue lines show estimates based on direct outcome modeling, ignoring unmeasured confounding,
and the gray bands represent 50\% and
95\% credible intervals based on our sensitivity analysis.
Importantly, because these results are derived from synthetic datasets, we can also compute the true policy outcomes,
which are indicated in the plot by the red lines.
Across the three levels of unmeasured confounding that we consider,
our sensitivity bands cover the ground truth line.
Moreover, the bands accurately reflect the true level of confounding, with wider bands when $x_i'$ is comprised of age, in which case there is substantial confounding,
and narrower bands when $x_i'$ is comprised of age, gender, and prior number of missed court appearance,
in which case there is relatively little confounding.

Finally, in Figure~\ref{fig:synthetic-het-tx}, we evaluate the performance of our sensitivity analysis method
at estimating subgroup treatment effects in the synthetic data.
As in our analysis of the real data
(Figure~\ref{fig:fulldata-het-tx}),
we consider subgroups defined by
age, gender,
and number of previously missed court appearances.
Each row in the plot corresponds to varying levels of censoring, and hence unmeasured confounding.
In each panel, the direct outcome modeling estimates---which assume there is no unmeasured confounding---are marked with \texttt{X}'s;
the thick and thin gray bars indicate 50\% and 95\% credible intervals, estimated using our sensitivity analysis approach; and the red points show the true subgroup treatment effects, calculated using the uncensored synthetic dataset $\Omega$.
Across conditions, the sensitivity bars capture the ground truth treatment effects for almost all subgroups.

\subsection{Comparison with previous methods for sensitivity analysis}

Sensitivity analysis %
for offline policy evaluation is
relatively new---we know of only one previous paper, \citet{simplerules}.
There are, however, several methods for assessing the sensitivity of average treatment effects to unmeasured confounding, which, as noted above, is a specific case of policy evaluation.
Here we compare our approach
to \citeauthor{simplerules}'s sensitivity method for offline policy evaluation,
as well as to two recent methods
designed for average treatment effects~\citep{mccandless2017comparison, dorie2016flexible}.

\begin{figure*}[t]
  \centering
  \includegraphics[width=14cm]{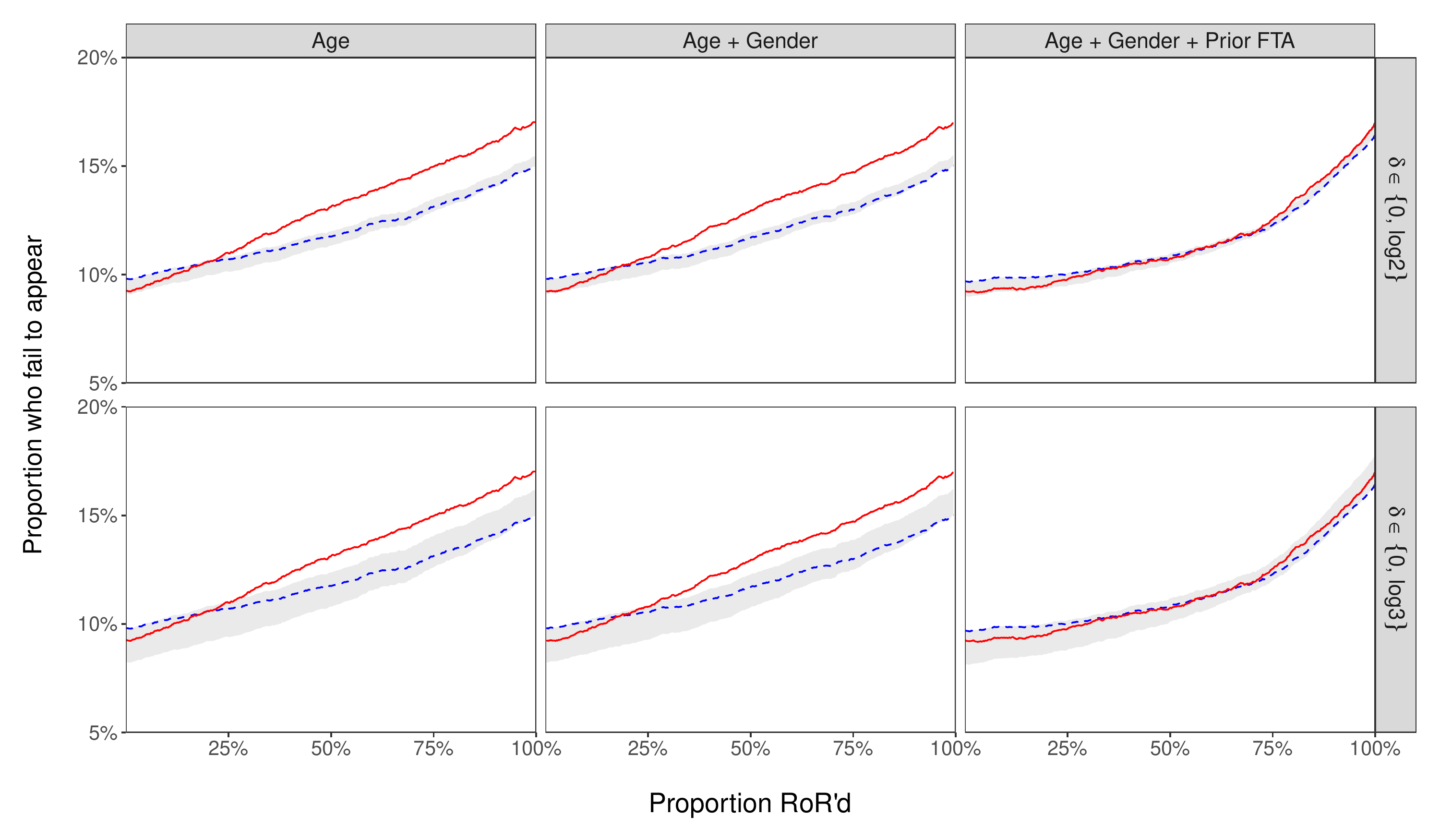}
  \caption{\emph{
  Sensitivity estimates based on the method of \citet{simplerules}, applied to three synthetic datasets with varying levels of confounding.
  The gray bands indicate minimum and maximum policy estimates for two parameter regimes:
  in the top row, the unobserved confounder can both double the odds of bail and of failing to appear at trial;
  in the bottom row, the unobserved confounder can triple these odds.
  The blue lines show estimates from direct outcome models ignoring unmeasured confounding, and the red lines indicate ground truth outcomes.
    }}
\label{fig:rnr-synthetic}
\end{figure*}

We start by comparing to the approach recently proposed by \cite{simplerules},
which adapts the classical method of
 \citet{rosenbaum1983assessing},
 to the setting of offline policy evaluation.
 As discussed in Section~\ref{sec:related_work}
 the \citeauthor{rosenbaum1983assessing} framework---and by extension the method of \citeauthor{simplerules}---assumes that there is
an unobserved
binary covariate $u$ that affects both treatment assignment and
expected response conditional on treatment.
For example, $u$ might indicate whether a defendant is particularly risky after accounting for factors recorded in the data.
There are four key parameters in this setup, which for concreteness we describe in the judicial context:
(1) $\Pr(u = 1)$, the prevalence of the unmeasured confounder;
(2) the effect of $u$ on a judge's decision to set bail or RoR a defendant;
(3) the effect of $u$ on the defendant's likelihood to miss court if RoR'd;
and (4) the effect of $u$ on the defendant's likelihood to miss court if bail is set.
By positing the magnitude of these effects
(e.g., based on prior domain knowledge),
one can estimate a plausible range of policy outcomes that account for unmeasured confounding.

Figure~\ref{fig:rnr-synthetic} shows the results of
the \citeauthor{simplerules}\ procedure applied to the three synthetic datasets described in Section~\ref{sec:simulation}.
In particular,
we compute the minimum and maximum policy estimates obtained by sweeping over two parameters regimes
suggested in their paper.
In the first regime (top),
we allow $\Pr(u = 1)$ to vary from 0.1 to 0.9,
assume that a defendant with $u=1$ has up to twice the odds of being detained as one with $u=0$,
and that $u$ can double the odds a defendant fails to appear, both if RoR'd or required to pay bail.
We also consider a more extreme situation (bottom),
where a defendant with $u=1$ has up to three times the odds of being detained as one with $u=0$,
and where $u$ can triple the odds a defendant fails to appear.
In each scenario, the red lines indicate the true outcomes of each policy, computed from the uncensored synthetic dataset,
the blue lines show estimates based on direct outcome models ignoring unmeasured confounding,
and the gray bands indicate the minimum and
maximum values of each policy over all parameter settings in the corresponding regime.

In contrast to our own sensitivity analysis results in Figure~\ref{fig:synthetic-sensitivity},
the sensitivity bands from the
\citeauthor{simplerules}\ approach
in Figure~\ref{fig:rnr-synthetic} often fail to capture the ground truth policy estimates.
Further, and more importantly,
the sensitivity bands do not appropriately adapt to the differing
levels of unmeasured confounding across datasets,
as indicated by their relatively constant
width across settings.
As a result, one must manually tune the sensitivity parameters
for each dataset to achieve satisfactory performance.
While not impossible---and indeed such calibration is the norm in classical sensitivity methods---the need for manual adjustment is a significant limitation of non-Bayesian approaches to sensitivity analysis.

Aside from the work of \citet{simplerules} discussed above, we know of no other off-the-shelf approaches for assessing the sensitivity of policy outcomes to unmeasured confounding.
However, as shown in Section~\ref{ssec:problem_def},
policy evaluation is a generalization of estimating average treatment effects,
and so we can compare our approach to methods designed for that problem.
In our judicial application, the average treatment effect is the difference in the proportion of defendants who fail to appear at court
when all are required to pay bail versus all being RoR'd.
We note that it is a particularly strong
test of our method to compare to approaches designed specifically to estimate the average treatment effect,
as our method is intended to address the much more general problem of policy evaluation.
In particular, one might expect that there is some cost to generalization, with our method exhibiting relatively worse performance on the narrow problem of estimating the average treatment effect in exchange for broader applicability.

\begin{figure*}[t]
  \centering
  \includegraphics[width=.8\linewidth]{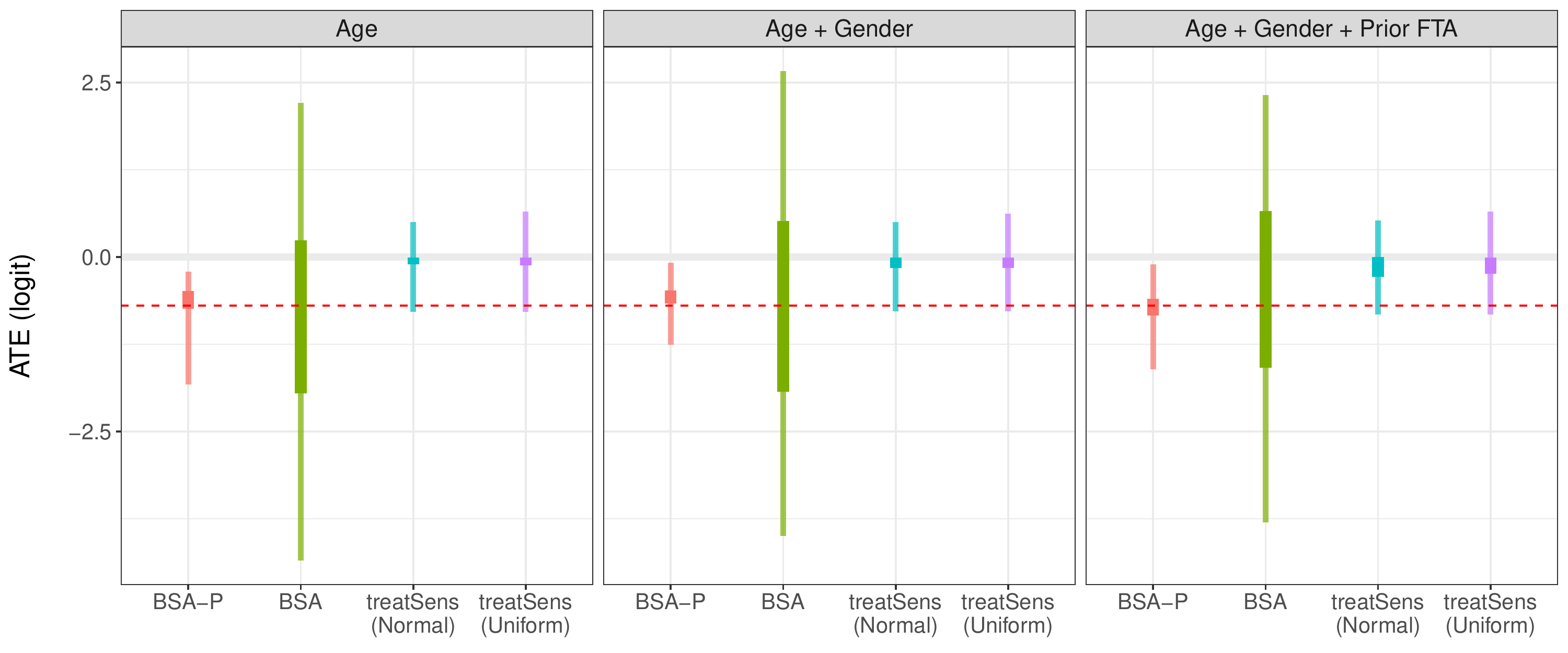}
  \caption{\emph{
  Estimates of average treatment effects for three synthetic datasets with differing levels of unmeasured confounded, as estimated by various methods:
  our own approach, labeled BSA-P;
  the method of~\citet{mccandless2017comparison},
  labeled BSA;
  and the method of~\citet{dorie2016flexible}, labeled TreatSens, for two different prior distributions.
    The think and thin lines represent 50\% and 95\% credible intervals, and the red horizontal line
    marks the true effect.}}
\label{fig:comparison-linerange}
\end{figure*}

We specifically compare our approach to two recently proposed
methods for Bayesian sensitivity analysis of average treatment effects:
the method of \citet{mccandless2017comparison}, which we refer to as BSA,
and the TreatSens method of \citet{dorie2016flexible}.
For space, we omit the details of these methods, but note that we implement the two sensitivity analysis procedures exactly as they were originally described;
in the case of TreatSens, we use the authors' publicly available \texttt{R} package to carry out our analysis.
Figure~\ref{fig:comparison-linerange} shows the results of estimating average treatment effects
on our three synthetic datasets described above,
where the true answer is indicated by the dashed red line,
and our approach is labeled BSA-P.
For TreatSens, since we can specify different priors on the unobservable $u$,
we compute results with a standard normal prior---as used in our own method---and with uniform priors, as suggested
by~\citeauthor{dorie2016flexible}
The thick and thin lines show the 50\% and 95\% credible intervals.
In all three synthetic datasets, representing different levels of unmeasured confounding, our approach is competitive with, and arguably even better than, the two methods we compare to.
In particular, whereas the true answer is at the periphery of the 95\% credible intervals generated by TreatSens, the ground truth lies near the posterior median of our approach.
Further, our credible intervals are substantially more narrow than those resulting from BSA, indicating that our method can simultaneously achieve accuracy and precision.

\section{Constructing optimal policies}
\label{sec:optimal}

In our analysis of judicial decisions,
we constructed a family of policies $\{\pi_s\}$ by
first estimating each defendant's flight risk
$\mu_0(x)$,
and then, for any threshold $s$,
defining the policy $\pi_s$ to recommend release
if and only if $\hat{\mu}_0(x) \leq s$.
This family of threshold policies is, necessarily, based
on a defendant's \emph{estimated} risk $\hat{\mu}_0(x)$---not his true risk $\mu_0(x)$---and as such may be affected by unmeasured confounding.
Here we explore the effects of such unmeasured confounding
on policy construction.
We first show, by example, that unmeasured confounding can in theory lead to pathological outcomes, with the riskiest defendants released and the least risky detained.
We then argue that such pathologies may not pose a substantial problem in practice.

To illustrate the potentially extreme effects of
unmeasured confounding on constructing policies,
we consider a simple, stylized example in Table~\ref{tab:policy}.
Suppose we only observe a defendant's age $x$,
which for simplicity we assume is either `young' or `old'.
In this example, and as indicated in  Table~\ref{tab:policy},
young defendants are, on average, riskier than old defendants; i.e., $\mu_0(\text{young}) > \mu_0(\text{old})$.
Based on this true risk ranking, one might reasonably
release old defendants while demanding bail from young ones.
Assume, however, that judges have access to additional information, and instead base their decisions on both a defendant's age $x$ and gender $u$, which we assume judges observe but we do not.
In our example, young men are the highest-risk group,
and so we further assume that judges
demand bail from young men while releasing
everyone else.
Now, because the observed flight risk for released young defendants
is relatively low---as
described in Table~\ref{tab:policy}---we
have that
$\hat{\mu}_0(\text{old}) > \hat{\mu}_0(\text{young})$,
the opposite of the true ranking.
Among \emph{released} defendants,
young defendants are indeed lower risk than old defendants,
but, importantly, this pattern does not hold in the full population.
Akin to Simpson's paradox, unmeasured confounding can cause one to infer a risk ranking of defendants that inverts the true order.
As such, one might mistakenly think it is reasonable to release young defendants (who, in the example, are actually high risk) while demanding bail from old defendants (who are low risk).

Despite this extreme hypothetical,
in practice, unobserved confounding may not significantly impact the policies one constructs.
Suppose, for example, that $\hat{\mu}_0(x)$ systematically underestimates each individual's risk by a constant factor.
In this case, even though the risk estimates are biased,
the policy
that recommends detaining the 10\% most risky defendants
is unaffected.
Particularly when faced with capacity constraints in allocating resources,
it often makes sense to parameterize policies in terms of risk quantiles (e.g., the proportion $p$ who are detained), rather than absolute risk thresholds (e.g., detaining those with risk exceeding $s$).
Fortunately, as we show next, risk rankings---and hence quantile-based policies---are relatively robust to unmeasured confounding.

We start by describing a relatively general setting in
which unmeasured confounding has no effect on risk rankings.
Though only an approximation of real-world conditions, this example provides insight into the problem.
Suppose that defendants are partitioned into discrete, observable groups $g_1, \dots, g_k$.
These groups might, for example, correspond to subsets defined by age, gender, criminal history, and other factors.
For illustration, we assume
that the true distribution of risk (i.e., a defendant's log-odds of failing to appear if released) within each group $g_i$ is
$\N(\theta_i, \sigma^2)$.
Importantly, as we discuss in more detail below,
we assume that the within-group variance $\sigma^2$ is constant across groups, though we allow the group means $\theta_i$ to differ.
We further assume that judges observe this true risk, and release defendants if and only if their (true) risk is below some fixed threshold $s$.
By definition,
group $g_i$ is, on average, riskier than group $g_j$
if and only if $\theta_i > \theta_j$.
The true, group-level risk ranking thus orders groups by $\theta_i$.
The problem is that we do not observe outcomes for a representative sample of defendants in each group---we only observe outcomes for those defendants who happened to be released---and so we cannot directly estimate $\theta_i$.
In this setup, however,
we can still recover the true risk ranking.
As we show in
Proposition~\ref{thm:monotonicity} below,
the mean of a right-truncated normal distribution is increasing in the
mean of the underlying normal distribution.
As a result,
even though risk estimates based on released defendants are biased,
these estimates preserve the true, group-level risk ranking.

\begin{table}[t]
	\caption{\emph{
An example illustrating that unmeasured confounding can lead to pathological policies.
Assume that young women, old women, and old men are RoR'd (the shaded rows),
    but only age is observed.
    If estimates of $\mu_0 = \Pr(\textrm{FTA}\mid \textrm{RoR})$ are based on those actually released,
    we will conclude the flight risk if RoR'd is $10\%$ for old defendants and
    $5\%$ for young defendants,
    and so old defendants would be deemed riskier than young defendants.
    However, while old individuals do have a $10\%$ flight risk,
    the true flight risk of young individuals if released is $(0.4*0.2 + 0.1*0.05)/0.5 = 17\%$, and so young defendants are in reality riskier than old ones in this case.
    }}
	\setlength{\tabcolsep}{5mm}
    \centering%
    \begin{tabular*}{13.58cm}{ccccccc}
      \toprule%
      $x$ & $u$ & Proportion & $t$ &
      $\mu_0(x, u)$ & $\hat{\mu}_0(x)$ & $\mu_0(x)$ \\
      \midrule%
      Young & M & 0.4 & 1 & 0.2 & & 0.17 \\
      \rowcolor{gray!25}
      Young & W & 0.1 & 0 & 0.05 & 0.05 & 0.17\\
      \rowcolor{gray!25}
      Old & M & 0.4 & 0 & 0.1 & 0.1 & 0.1\\
      \rowcolor{gray!25}
      Old & W & 0.1 & 0 & 0.1 & 0.1 & 0.1\\
      \bottomrule%
    \end{tabular*}
    \label{tab:policy}
\end{table}

\begin{proposition}
\label{thm:monotonicity}
  Let $R_i \sim \mathrm{N}(\theta_i, \sigma^2)$, then if $\theta_1 < \theta_2$, we have $\EE(R_1 \mid R_1 < s) < \EE(R_2 \mid R_2 < s)$.  That is, for fixed $s$ and $\sigma^2$, the mean of the right-truncated normal distribution is an increasing function of the mean of the underlying normal distribution.
\end{proposition}

\begin{figure*}[t]
  \centering
  \includegraphics[width=14cm]{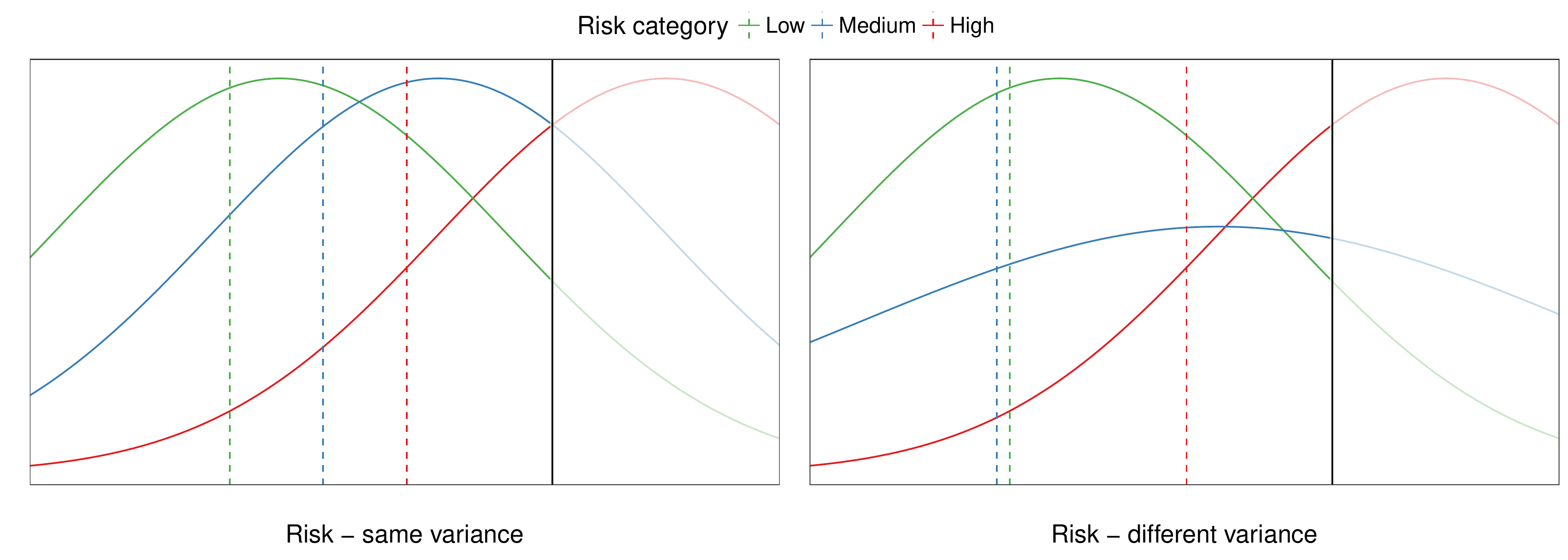}
  \caption{\emph{Illustration of the ordering of means of truncated normal distributions.
    The solid green, blue, and red curves represent the complete risk distribution for
    low, medium, and high risk subgroups, respectively.
    The vertical black line is the truncation value;
    the threshold at which a release decision is made.
    Dashed vertical lines represent means of the same distributions,
    conditioned on being below the threshold.
    If the variance of each distribution is identical (left),
    the ordering of the means of the truncated distributions is the same as the
    ordering of the means of the original, non-truncated distributions.
    When the variances differ (right),
    the these two orderings may also diverge.
    }}
\label{fig:truncated-normal}
\end{figure*}

A proof of Proposition~\ref{thm:monotonicity}
is given in the Appendix.
We present a visual representation of the result
in the left panel of Figure~\ref{fig:truncated-normal}.
Based only on the means of the truncated distributions,
one can recover the true group ranking.
The key assumption in Proposition~\ref{thm:monotonicity} is
that the within-group variance $\sigma^2$ is constant across groups.
Without this assumption,
one cannot always recover risk rankings,
as illustrated in the
right panel of Figure~\ref{fig:truncated-normal}.
In that example, the medium-risk group (shown in blue)
has higher variance than the low risk group (shown in green), distorting the ranking inferred from the truncated distributions.
A similar phenomenon is at play in the example described in
Table~\ref{tab:policy}.
In that case, the within-group variance of young defendants is substantially higher than for old defendants,
corrupting the risk ranking.
In both examples, one requires quite large differences in within-group variance to alter the risk ranking,
pointing to the robustness of such rankings in the presence of unmeasured confounding.

Proposition~\ref{thm:monotonicity}
establishes one particular set of assumptions under which
risk rankings can be recovered in the presence of unmeasured confounding.
We now present empirical evidence that
unmeasured confounding does not substantially impact the policies one learns in our bail application.
As in Section~\ref{sec:simulation},
we construct a synthetic dataset
in which we have access to both
potential outcomes for each defendant: failure to appear if RoR'd $y_i(0)$, and failure to appear if bail is required $y_i(1)$.
As before, we censor the data by restricting to three different subsets of covariates: (1) age; (2) age and gender; and (3)
age, gender, and number of previously missed court dates.
For each subset of covariates, we derive two families of policies by ranking defendants along the observed dimensions.
The first set of policies is based only on outcomes for released defendants.
That is, as we would do in practice, we restrict to released defendants and fit a model that estimates flight risk conditional on the observed covariates. This fitted model is then used to estimate the flight risk of---and subsequently rank---all defendants, both those who were released and those who were not.
The second set of policies is based on information for
all defendants, where we use the counterfactual outcome $y_i(0)$ for those defendants who were not released.
In this case, the derived ranking is the best one can do with the available covariates.
Of course, such an optimal policy cannot generally be learned in practice, but it is a natural benchmark to consider when assessing the effects of unmeasured confounding on policy construction.

\begin{figure*}[t]
  \centering
  \includegraphics[width=14cm]{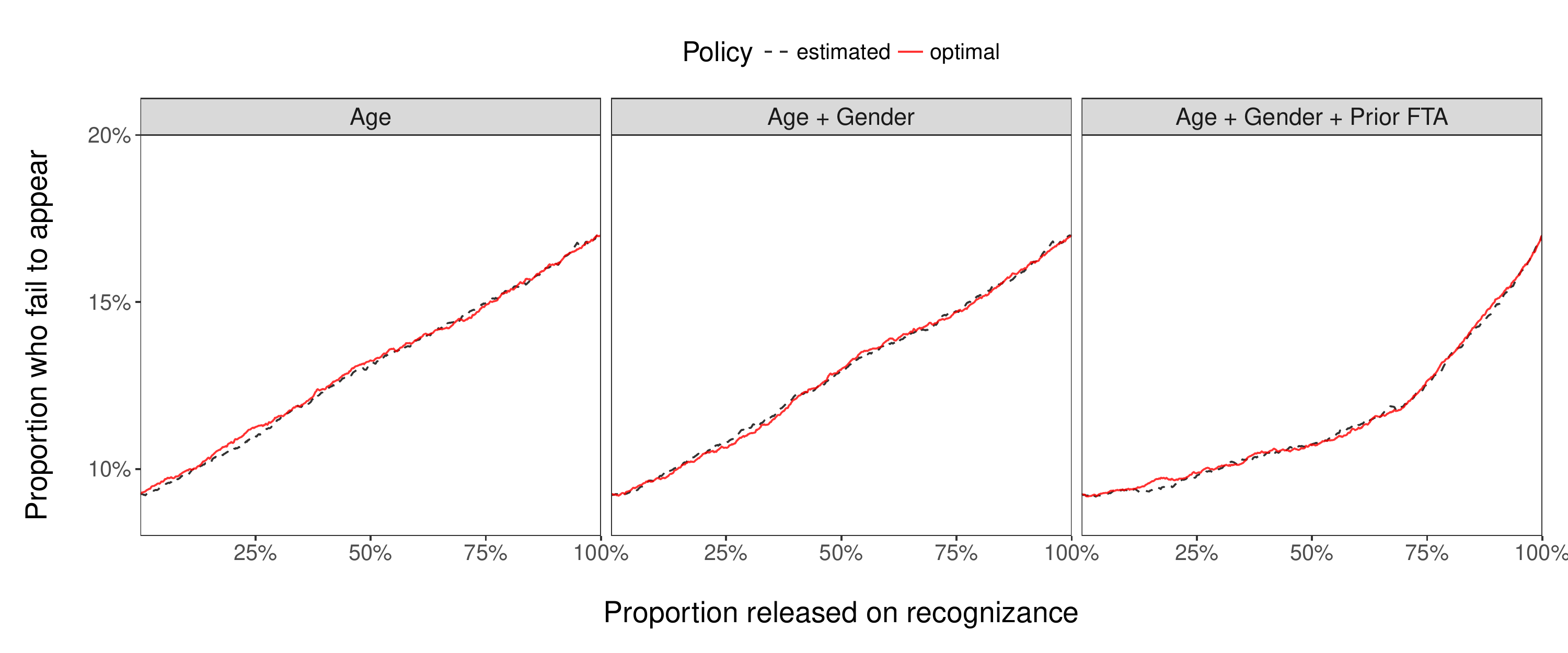}
  \caption{\emph{Policies constructed from the observed outcomes (black dashed lines) are nearly identical to the optimal policies (red lines).
  In the former case, risk rankings are estimated based on outcomes for the subset of released defendants; in the latter case, they are constructed with information on the full set of defendants, using the potential outcome $y_i(0)$ even if the defendant was not released.
    }}
\label{fig:synthetic-policy}
\end{figure*}

In Figure~\ref{fig:synthetic-policy},
we evaluate both families of policies described above, comparing the proportion who are released to the proportion that fail to appear under each.
The proportion who fail to appear is computed using the ground truth counterfactuals $y_i(0)$ and $y_i(1)$.
The red lines show results for the optimal policies, derived based on complete information about the potential outcomes.
The black dashed lines show results for the policies one would learn in practice, by considering only the outcomes of those who were released.
In all three conditions, the black and red lines are almost identical.
Thus, even though unmeasured confounding can in theory lead to suboptimal risk rankings,
it appears, at least in our case, that policy construction is robust to such hidden heterogeneity.

\section{Discussion}
As algorithms are increasingly used to guide high-stakes decisions,
it has become ever more important to accurately assess their likely effects.
Here we have addressed this problem of offline policy evaluation by coupling classical statistical ideas on sensitivity analysis with modern methods of machine learning and large-scale Bayesian inference.
The result is a conceptually simple yet powerful technique for evaluating algorithmic decision rules.

One of the key strengths of our general strategy is its flexibility.
In our bail application, both the treatment (RoR or bail) and the response (FTA or not) are binary.
In principle, however, our approach can handle much more complicated treatment choices and outcomes simply by specifying alternative model forms.
For example, instead of modeling the bail decision as binary, one might model treatment as a multinomial or continuous function corresponding to the amount of bail required of the defendant.
Such modeling changes create relatively little overhead for practitioners, since one can use black-box Bayesian inference to compute the posterior, circumventing the difficult analytic calculations that have hindered past methods.
More complicated models typically come with longer inference times, though advances in Bayesian computation---including black-box variational inference~\citep{ranganath2014}---have produced algorithms capable of handling substantial complexity.

By definition, it is impossible to precisely quantify \emph{unmeasured} confounding, and so all methods of sensitivity analysis require  assumptions that are inherently untestable.
Traditional methods handle this situation by
requiring practitioners to specify parameters describing the structure and scale of the assumed confounding, informed, for example, by domain expertise.
In contrast to that traditional approach,
our Bayesian framework largely obviates the need to explicitly set sensitivity parameters.
But there is no free lunch.
In our case, the necessary side information enters through both the assumed form of the data generating process and the priors.
By adopting an expressive model form and setting weakly informative priors, our approach balances the need to provide at least some information about the structure of the potential confounding with the impossibility of specifying it exactly.
This middle ground appears to work well in practice,
but it is useful to remember the conceptual underpinnings of our strategy when applying it to new domains.

Over the last two decades, sophisticated methods of machine learning have emerged and gained widespread adoption.
More recently, these methods have been applied to traditional problems of causal inference and their modern incarnations, like offline policy evaluation.
Machine learning and causal inference are two sides of the same coin, but the links between the two are still under-developed.
Here we have bridged one such gap by porting ideas from classical sensitivity analysis to algorithmic decision making.
Looking forward, we hope that sensitivity analysis is more tightly integrated into machine learning pipelines
and, more generally, that our work spurs further connections between methods of causal inference and prediction.

\section*{Acknowledgements}
We thank Alexander D'Amour and Andrew Gelman for helpful comments.

\bibliography{icml2017}
\bibliographystyle{apalike}

\pagebreak
\appendix
\appendixpage

\section{Additional model details}

In the main text, we described the likelihood function for our Bayesian model of unmeasured confounding.
Here we complete the model specification
by describing the prior distribution on the parameters.

On each of the unmeasured confounders $u_i$, we set a $\N(0,1)$ prior.
On the coefficients $\alpha$, $\beta$, and $\gamma$ we use a random-walk prior.
Intuitively, random-walk priors ensure that adjacent groups have similar coefficient values, mitigating the dependence of results on the exact number of groups $K$.
Formally, the random-walk prior on $\alpha_0$ is given by
\begin{align*}
\alpha_{0,1} & \sim \N(0, 1) \\
\alpha_{0,j} & \sim \N(\alpha_{0,j-1}, \tau^2_{\alpha_0}) \quad \text{for} \ j \in \{2, 3, \dots, K\} \\
\tau_{\alpha_0} & \sim \N_+(0, \sigma_\tau^2),
\end{align*}
where $\N_+(0, \sigma_\tau^2)$ indicates the half-normal distribution with standard deviation $\sigma_\tau$.
We analogously set priors for the parameters
$\alpha_{\hat{\mu}_0}$,
$\beta_0$,
$\beta_{\hat{\mu}_1}$,
$\gamma_0$,
and $\gamma_{\hat{e}}$.

\begin{figure*}[t]
  \centering
  \includegraphics[width=14cm]{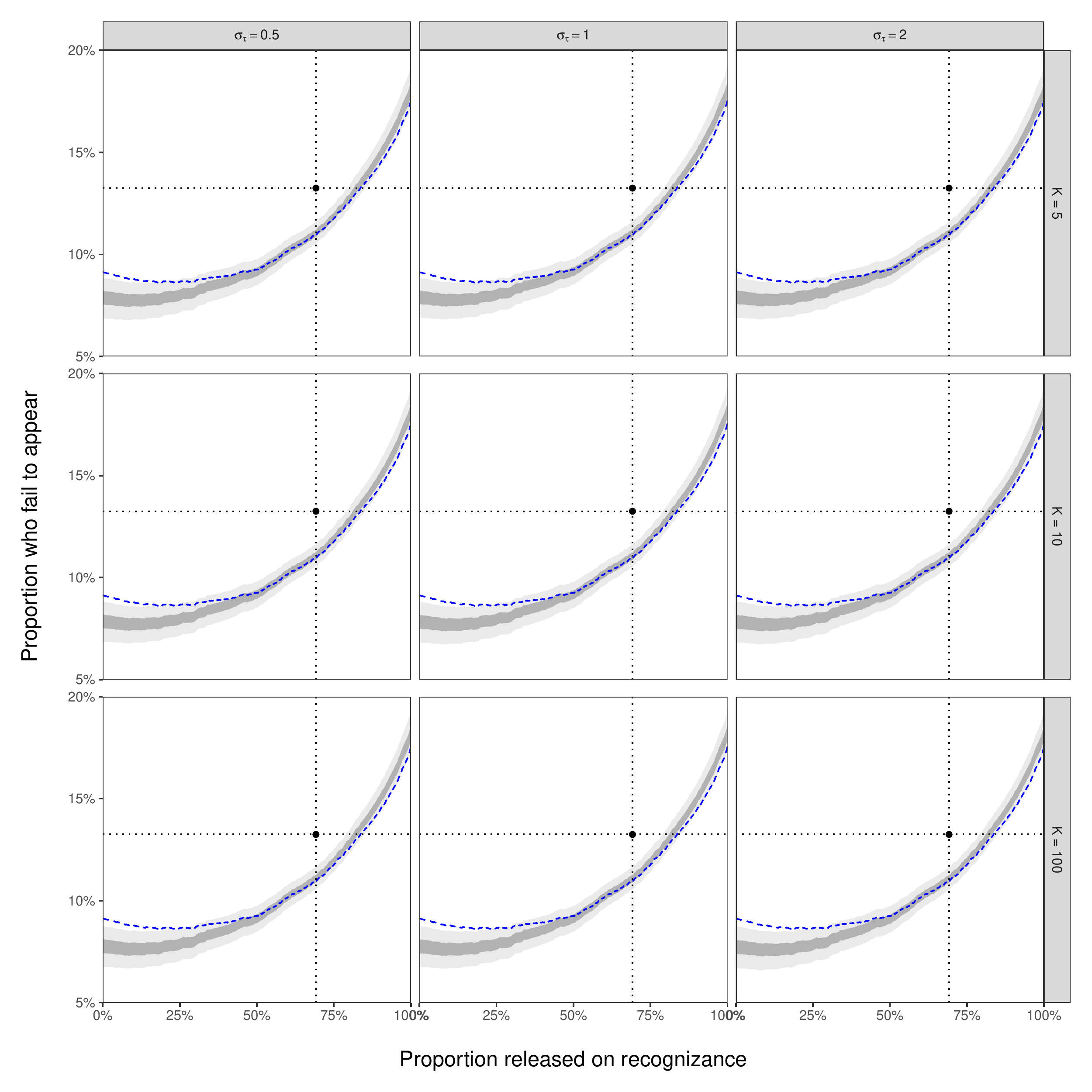}
  \caption{\emph{Main result of Figure~\ref{fig:fulldata-sensitivity}, repeated for different values of the parameter $K$ (rows) and prior distribution, parameterized with $\sigma_\tau$ (columns).
  The results in each panel are nearly identical, indicating that they are not substantially affected by the particular choice of $K$ or $\sigma_\tau$ in this context.
    }}
\label{fig:fulldata-insensitive}
\end{figure*}

For the coefficients on the unobserved confounders $u_i$, we set random-walk priors with an additional constraint to ensure positivity,
so that $\Pr(T=1)$, $\Pr(Y(0)=1)$, and
$\Pr(Y(1)=1)$ all increase with $u_i$.
In the judicial context, for example, one can imagine that $u_i$ corresponds to unobserved risk of failing to appear, with judges more likely to demand bail from riskier defendants. This constraint, while not strictly necessary, facilitates estimation of the posterior distribution. Formally, for $\gamma_u$ we have
\begin{align*}
\alpha_{u,1} & \sim \N_+(0, 1) \\
\alpha_{u,j} & \sim \N_+(\alpha_{u,j-1}, \tau^2_{\alpha_u}) \quad \text{for} \ j \in \{2, 3, \dots, K\} \\
\tau_{\alpha_u} & \sim \N_+(0, \sigma_\tau^2).
\end{align*}
We similarly set sign-constrained random-walk priors on $\beta_u$ and $\gamma_u$.
For the main results in this paper, we set values of $\sigma_\tau = 1$ and $K = 10$.
However, Figure~\ref{fig:fulldata-insensitive} shows that the results are not substantially affected by the choice of $K$ or the prior distribution (parameterized by $\sigma_{\tau}$).

\section{Proof of Proposition~\ref{thm:monotonicity}}

  As is well-known, if a random variable $R$ is normally distributed with mean $\theta$ and variance $\sigma^2$, the mean and variance of the truncated variable $(R\mid R < s)$ are given by
  $$\EE(R\mid R < s) = \theta - \sigma \frac{\phi(\beta)}{\Phi(\beta)}$$
  and
  $$\Var(R\mid R < s) = \sigma^2 \left[1 - \beta\frac{\phi(\beta)}{\Phi(\beta)}
    - \left(\frac{\phi(\beta)}{\Phi(\beta)} \right)^2 \right],$$
  where $\phi$ and $\Phi$ are the PDF and CDF of the standard normal distribution, respectively,
  and $\beta = (s - \theta)/\sigma$.
  Differentiating $\EE(R\mid R < s)$ with respect to $\theta$, and using the identities $\frac{d\phi}{d\theta} = \frac{\beta\phi(\beta)}{\sigma}$ and $\frac{d\Phi}{d\theta} = -\frac{\phi(\beta)}{\sigma}$, we see that
  \begin{align*}
    \frac{d\EE(R\mid R < s)}{d\theta} & = 1 -
      \beta\frac{\phi(\beta)}{\Phi(\beta)}  - \left(\frac{\phi(\beta)}{\Phi(\beta)} \right)^2 \\
      & = \frac{\Var(R\mid R < s)}{\sigma^2}. \\
  \end{align*}
  Since this derivative is the ratio of two variances, it is positive, hence $\EE(R\mid R < s)$ is increasing in $\theta$, as desired.
$\qed$

\end{document}